\documentstyle[sprocl]{article}

\def\be{\begin{equation}}
\def\ee{\end{equation}}
\def\bea{\begin{eqnarray}}
\def\eea{\end{eqnarray}}
\begin{document}

\title{BEYOND HTL: THE CLASSICAL KINETIC THEORY \\
OF LANDAU DAMPING\\ FOR SELFINTERACTING SCALAR FIELDS\\
IN THE BROKEN PHASE}

\author{Andr\'as Patk\'os, Zsolt Sz\'ep}

\address{Department of Atomic Physics, E\"otv\"os University, P\'azm\'any 
P\'eter s\'et\'any 1/A\\ 
H-1117 Hungary\\E-mail: patkos@ludens.elte.hu, szepzs@cleopatra.elte.hu} 


\maketitle\abstracts{The effective theory of low frequency fluctuations of  
selfinteracting scalar fields is constructed in the broken symmetry phase. 
The theory resulting from integrating
fluctuations with frequencies much
above the spontanously generated mass scale $(p_0>>M)$
is found to be local. Non-local dynamics, especially Landau damping emerges 
under the effect of fluctuations in the 
$p_0 \sim M$ region. A kinetic theory of relativistic scalar gas particles 
interacting via their locally variable mass with the low 
frequency scalar field is shown to be
equivalent to this effective field theory for scales below the 
characteristic mass, that is 
beyond the accuracy of the Hard Thermal Loop (HTL) approximation.}

\section{Effective equation for the slow modes}
The model investigated in this paper has the Lagrangian density
\be
L={1\over 2}(\partial_\mu\varphi(x))^2-{1\over 2}m^2\varphi^2(x)-
{\lambda\over 24}\varphi^4(x).
\label{Ldensity}
\ee
The field $\varphi(x)$ is split into the sum of terms with low 
($p_0<\Lambda$)
and high ($p_0>\Lambda$) frequency Fourier-components, that is $
\varphi (x)=\tilde\Phi (x)+\phi (x), \langle \phi(x)\rangle =0$.
The classical equation of motion for the low frequency component $\tilde\Phi 
(x)$ is the following:
\be
(\partial^2+m^2)\tilde\Phi (x)+{\lambda\over 6}[\tilde\Phi^3(x)+3\tilde\Phi 
(x)\phi^2(x)+3\tilde\Phi^2(x)\phi (x)+\phi^3(x)]=0.
\ee
The effective equation of motion arises upon averaging over the (quantum) 
fluctuations of the high frequency field $\phi(x)$:
\be
(\partial^2+m^2)\tilde\Phi (x)+{\lambda\over 6}[\tilde\Phi^3(x)+3\tilde\Phi 
(x)\langle\phi^2(x)\rangle ] =0.
\ee
The last term on the left hand side represents the source induced by the
action of the high frequency modes. It is a functional of $\tilde\Phi (x)$.
The action of the effective field theory may be reconstructed from this
equation. This approach is closely related to the Thermal Renormalisation
Group equation of D'Attanasio and Pietroni \cite{AP96}.

In the broken phase the non-zero average value spontanously generated below
$T_c$ is separated from the low-frequency part, $
\tilde\Phi (x)=\bar\Phi +\Phi (x)$.
The expectation value $\bar\Phi$ is determined by the effective equation
\be
m^2+{\lambda\over 2}\langle\phi^2(x)\rangle^{(0)}+{\lambda\over 6}\bar\Phi^2=0
\label{effvev},
\ee
where we have introduced the indexed expectation value
\be
\langle\phi^2(x)\rangle^{(j)}\sim \Phi^j
\ee
to denote that part of the full expectation value which is "proportional" 
functionally to the j-th power of $\Phi (x)$. 

Our present goal is to determine the effective {\it linear} dynamics of the 
$\Phi$-field, therefore it is sufficient to study the linearised effective
equation for $\Phi (x)$:
\be
(\partial^2+{\lambda\over 3}\bar\Phi^2)\Phi (x)=-{\lambda\over 2}\bar\Phi
\langle\phi^2(x)\rangle^{(1)}.
\ee
(In this equation $\bar\Phi$ is the solution of (\ref{effvev})). Clearly, 
the linear response theory of (\ref{Ldensity}) is contained in the 
induced current, determined by $\langle\phi^2(x)\rangle^{(1)}$.
 
\section{Statistics of the high frequency modes}
For the computation of the leading effect of the high-frequency modes in
the low frequency projection of the equation of motion it is sufficient
to study the  two-point function $\langle\phi (x)\phi (y)\rangle$.
For its determination we follow the procedure carefully described by 
Mr\'owczy\'nski and Danielewicz \cite{MR90}.

Introducing the Wigner transform by the relation
\be
\langle\phi (x)\phi (y)\rangle =\int {d^4p\over (2\pi )^4}e^{-ip(x-y)}\Delta
(X,p), \qquad X={x+y\over 2}.
\ee
one arrives at the following equations for $\Delta (X,p)$ 
\bea
&
({1\over 4}\partial_X^2-p^2+M^2(X))\Delta (X,p)=0,\nonumber\\
&
(p\partial_X+{1\over 2}\partial_XM^2(X)\partial_p)\Delta (X,p)=0,
\label{MDeq}
\eea
where  $M^2(x)=m^2+\lambda\tilde\Phi^2(x)/2$.
The quantity appearing in the induced source is related to the 
Wigner transform of the two-point function through the relation
\be
\langle\phi^2(x)\rangle^{(1)}=\int{d^4p\over (2\pi )^4}\Delta^{(1)} (x=X,p).
\ee

The important limitation on the range of validity of the effective dynamics 
is expressed by the assumption that the second derivative with respect 
to $X$ is negligible
relative to $p^2$ and $M^2$ on the left hand side of the first equation of
(\ref{MDeq}). Then this equation is transformed simply into a local mass-shell
condition, while the second equation of (\ref{MDeq}) can be interpreted
as a Boltzmann-equation for the phase-space ``distribution function'' 
$\Delta (X,p)$.

The background-independent solution $\Delta ^{(0)}$ is given by the 
well-known 
free correlator, slightly modified to account for the lower frequency cut
appearing in the Fourier series expansion of $\phi (x)$:
\be
\Delta^{(0)}(X,p)=(\Theta (p_0)+\tilde n(|p_0|))2\pi\delta (p^2-M^2),\quad
\tilde n(p_0)={1\over e^{\beta |p_0|}-1}\Theta (|p_0|-\Lambda).
\label{freecorr}
\ee
An iterated solution of the second equation of Eq.(\ref{MDeq}) starting from
 (\ref{freecorr}) yields
\be
\Delta^{(1)}(X,p)=-{\lambda\over 2}\bar\Phi(p\partial_X)^{-1}{\partial\Phi 
(X)\over \partial X_\mu}{\partial\Delta^{(0)}(X,p)\over \partial p_\mu}.
\label{delta1}
\ee

\section{The induced source}

For the analysis of the induced source
\be
j_{ind}(x)=-{\lambda\over 2}\bar\Phi\langle\phi^2(x)\rangle^{(1)}
={\lambda^2\bar\Phi^2\over 4}\int{d^4p\over (2\pi)^4}(p\partial_x)^{-1}
\partial_x\Phi\partial_p\Delta^{(0)}(x,p)
\ee
one takes its Fourier-transform with respect to $x$. Using the explicit 
expression (\ref{freecorr}) one easily recognizes the only non-trivial 
(non-local) contribution arises from the integral
\be
\int{d^4p\over (2\pi)^3}{1\over pk}k_0\delta (p^2-M^2){d\tilde n\over dp_0}.
\ee
Its imaginary part determines the rate of Landau-damping. Simple integration 
steps lead to 
\bea
&
{\rm Im}j(k_0,k)=-{\lambda^2\bar\Phi^2\over 16\pi}{k_0\over k}\Phi (k)\Theta
(k^2-k_0^2)\int_{t_0}^\infty dt{d\tilde n\over dt},\nonumber\\ 
&
\tilde n(t)={1\over e^{\beta Mt}-1}\Theta (Mt-\Lambda),\quad t_0=
{1/\sqrt{1-(k_0 /k)^2}}.
\eea
This integral is zero if $\Lambda >Mt_0$, but for $\Lambda <Mt_0$ it gives
\be
{\rm Im}j(k_0,k)={\lambda^2\bar\Phi^2\over 16\pi}{k_0\over k}\Phi (k)\Theta
(k^2-k_0^2){1\over e^{\beta M/\sqrt{1-(k_0 /k)^2}}-1}
\ee
independent of the value of $\Lambda$.

The result has very transparent interpretation. In the HTL-limit, when only 
 the modes with much higher frequencies than any mass scale 
in the theory are taken into account, no Landau-damping arises. The
effective theory is local! 

Going beyond HTL, ($k_0<<\Lambda << M$) the correct non-local
dynamics (reflected also by the Landau damping) originating from
the 1-loop self-energy contribution is recovered, when comparison with Boyanovsky {\it et al.} is made \cite{BO95}. 

\section{Classical mechanical representation of the non-local dynamics}

Recently we proposed to superimpose on the scalar field theory 
(\ref{Ldensity}) a gas
of relativistic scalar particles with the action \cite{SZP99}
\be
S_{mech}=-\sum_i\int d\tau M_{loc}[\bar\Phi,\Phi (\xi_i (\tau ))], \qquad
M_{loc}^2[\Phi (\xi_i)]=m^2+{\lambda\over 2}\tilde\Phi^2(\xi_i(\tau )),
\label{mechact}
\ee
where $\xi_i(\tau )$ denotes the world-line of the $i$-th particle
of the gas.
The mass of these particles varies with the field along the 
trajectory of the particles. The equation of motion of one of the
particles is given by
\be
M_{loc}(\xi ){dp_\mu\over d\tau}={1\over 2}
{\partial M_{loc}^2(\xi )\over \partial \xi_\mu}.
\ee
The kinetic equation for the collisionless evolution of the 
one-particle phase-space density $f(x,p)$ of this gas is
\be
p_\mu{\partial f(x,p)\over\partial x_\mu}+
M_{loc}{dp_\mu\over d\tau}{\partial \over \partial p_\mu}
f(x,p)=0,
\ee
which clearly agrees with the second equation of (\ref{MDeq}), while the 
solution of the first one has dictated the choice of the effective mass 
expression in (\ref{mechact}). 

Variation of (\ref{mechact}) with respect to $\Phi (x)$ leads to the 
term corresponding to the induced source density in the wave equation:
\be
j_{ind}={\lambda\over 2}\int{d^3p\over (2\pi )^3p_0}(\bar\Phi+\Phi(x))f(x,p)
\ee
$(p_0^2=p^2+M^2)$.
This expression is obtained by averaging the contributions of the different 
particle-trajectories in the gas, passing nearby the point $x$ with any 
momentum $p$:
\be
\langle\int d\tau\sum_i\delta^{(4)}(x-\xi_i(\tau ))\rangle
=M_{loc}\int{d^3p\over (2\pi)^3p_0}f(x,p).
\ee
Using the solution of the Boltzmann equation obtained upon iteration 
starting from the equilibrium Bose-Einstein factor,
one finds the same expression for the non-local part
of the source, as was found above implying the same result also for the
rate of Landau damping.

\section{Conclusion}
We have presented two equivalent methods of treating the effective
dynamics of the low-frequency fluctuations of self-interacting scalar fields
in the broken phase of the theory. The dynamics is proved to be non-local
if the effect of fluctuation modes below the mass scale $M$ is also taken 
into account. A fully local representation was proposed by superimposing
a relativistic gas with specially chosen field dependent mass on the
original field theory.

The range of validity of the fully local version
of the effective model can be ascertained only from its
comparison with the result of lowering the separation scale $\Lambda$
in the detailed integration
over the fluctuations with different frequencies. From the comparison
one learns that the combined kinetic plus field theory is equivalent
to the effective theory for the modes with $k_0<<\Lambda< M$, that is its
validity goes beyond the Hard Thermal Loop approximation.

\section*{Acknowledgements}

This work has benefited from discussions with Antal Jakov\'ac and 
P\'eter Petreczky. Helpful comments of U. Heinz and C. Manuel during SEWM 98 
are gratefully acknowledged. We thank the organizers the creative
atmosphere of the meeting.

\end{document}